\documentclass[english,11pt,a4paper]{article}
\usepackage[T1]{fontenc}
\usepackage[latin9]{inputenc}
\usepackage{booktabs}
\usepackage{amsmath}
\usepackage{amsthm}
\usepackage{amssymb}
\usepackage{graphicx}

\makeatletter

\providecommand{\tabularnewline}{\\}

\numberwithin{equation}{section}

\usepackage{jheppub}
\usepackage{slashed}

\makeatother

\usepackage{babel}
\begin{document}
\title{ $N_{{\rm eff}}$ constraints on light mediators coupled to
neutrinos: the dilution-resistant effect }

\abstract{ We investigate the impact of new light particles, carrying
significant energy in the early universe after neutrino decoupling,
on the cosmological effective relativistic neutrino species, $N_{{\rm eff}}$.
If the light particles are produced from decoupled neutrinos, $N_{{\rm eff}}$
is predominantly modified through the dilution-resistant effect. This
effect arises because the energy stored in the mass of new particles
is less diluted than the photon and neutrino energy as the universe
expands. Our study comprehensively explores this effect, deriving
$N_{{\rm eff}}$ constraints on the couplings of light mediators with
neutrinos, encompassing both scalar and vector mediators. We find
that the dilution-resistant effect can increase $N_{{\rm eff}}$ by
0.118 and 0.242 for scalar and vector mediators, respectively.  These
values can be readily reached by forthcoming CMB experiments. Upon
reaching these levels, future $N_{{\rm eff}}$ constraints on the
couplings will be improved by many orders of magnitude.

}

\author[a]{Shao-Ping Li} 
\author[a]{and Xun-Jie Xu} 
\affiliation[a]{Institute of High Energy Physics, Chinese Academy of Sciences, Beijing 100049, China} 
\preprint{\today}  
\emailAdd{spli@ihep.ac.cn} 
\emailAdd{xuxj@ihep.ac.cn}

\maketitle

\section{Introduction}

 In the era of precision measurements, modern cosmology has achieved
numerous excellent agreements between observations and theoretical
understandings.  For instance, the effective relativistic neutrino
species, $N_{{\rm eff}}$, has been precisely measured to be $N_{{\rm eff}}=2.99\pm0.17$~\cite{Planck:2018vyg},
exhibiting good agreement with the prediction of the standard model
(SM),  $N_{{\rm eff}}^{{\rm SM}}\approx3.045$~\cite{deSalas:2016ztq,EscuderoAbenza:2020cmq,Akita:2020szl,Bennett:2020zkv,Cielo:2023bqp}.
 Looking ahead, future precision measurements of $N_{{\rm eff}}$
at CMB-S4~\cite{Abazajian:2019eic,CMB-S4:2016ple}, SPT-3G~\cite{SPT-3G:2014dbx},
Simons Observatory~\cite{SimonsObservatory:2019qwx,SimonsObservatory:2018koc},
PICO~\cite{NASAPICO:2019thw}, CORE~\cite{CORE:2016npo} and CMB-HD~\cite{CMB-HD:2022bsz}
are anticipated to reach the percent level, providing an excellent
opportunity to thoroughly probe the SM prediction, including the small
deviation from three.

 Precision measurements of $N_{{\rm eff}}$ offer a promising avenue
to shed light on new physics beyond the SM, given that many new physics
scenarios predict significant deviations of $N_{{\rm eff}}$ from
the SM value~\cite{Boehm:2012gr,deSalas:2016ztq,Huang:2017egl,Borah:2018gjk,Escudero:2018mvt,Escudero:2019gzq,Abazajian:2019oqj,Depta:2019lbe,EscuderoAbenza:2020cmq,Luo:2020sho,Luo:2020fdt,Borah:2020boy,Adshead:2020ekg,Venzor:2020ova,Li:2021okx,Hufnagel:2021pso,Li:2022dkc,Ghosh:2022fws,Ganguly:2022ujt,Sandner:2023ptm,Ghosh:2023ocl}.
  For instance, Dirac neutrinos with thermalized right-handed components
would lead to $\Delta N_{{\rm eff}}\equiv N_{{\rm eff}}-N_{{\rm eff}}^{{\rm SM}}\geq0.14$~\cite{Abazajian:2019oqj,Luo:2020sho,Luo:2020fdt,Li:2022yna}.
 Axions or axion-like particles, if thermalized before the electroweak
phase transition, would  cause $\Delta N_{{\rm eff}}=0.027$~\cite{Baumann:2016wac}.
 Unstable particles may also leave observable imprints in $N_{{\rm eff}}$
if a considerable fraction of these particles decay after neutrino
decoupling. This has been used to set stringent constraints on light
mediators of new interactions~\cite{Kamada:2015era,Knapen:2017xzo,Kamada:2018zxi,Escudero:2019gzq}.

Generally speaking, a new species can modify $N_{{\rm eff}}$ if it
has been produced  before neutrino decoupling and carries a certain
amount of energy after neutrino decoupling.  If the energy carried
by the new species at the moment of neutrino decoupling is low (e.g.,
due to the Boltzmann suppression  or insufficient production), $\Delta N_{{\rm eff}}$
is expected to be small. 

However, we would like to emphasize here that even if this part of
energy is vanishingly small, new species produced after neutrino decoupling
might still cause observably large $\Delta N_{{\rm eff}}$ due to
the mass effect of the introduced new species. Consider for example
a new light scalar $\phi$ that is dominantly coupled to neutrinos
($\nu$), with the mass $m_{\phi}=1$ keV and the coupling $g_{\phi}=10^{-9}$.
Such a species remains unthermalized until the temperature drops down
to about 30 keV\footnote{This can be seen either from our Fig.~\ref{fig:approx} or from a
simple estimate using the thermalization condition $\langle\Gamma_{2\nu\to\phi}\rangle\gtrsim H$
where $\langle\Gamma_{2\nu\to\phi}\rangle\approx\frac{g_{\phi}^{2}m_{\phi}^{2}}{16\pi T_{\nu}}$
is the thermal average of the inverse decay rate and $H\approx6T_{\nu}^{2}/m_{{\rm pl}}$
is the Hubble expansion rate at the keV scale. It is straightforward
to see that $\langle\Gamma_{2\nu\to\phi}\rangle\gtrsim H$ requires
$T_{\nu}\lesssim34$ keV. }, and then starts to thermalize (i.e.~being substantially produced
from $\nu$). Eventually, all produced $\phi$ particles will decay
and release the energy back to $\nu$ at temperatures well below $m_{\phi}$.
So during the entire process, $\phi$ first absorbs energy from $\nu$
and then returns it to $\nu$. If the energy densities of $\phi$
and $\nu$ both scale as $a^{-4}$ where $a$ is the scale factor
of the expanding universe, the total energy in a comoving volume,
$\rho_{\phi+\nu}a^{4}$, should remain constant, implying that neutrinos
would not gain any energy from this process. However, since a significant
amount of the energy is stored in the form of $m_{\phi}$, which is
resistant to the dilution caused by the Hubble expansion, $\rho_{\phi+\nu}a^{4}$
actually increases during the process. We refer to this effect as
the dilution-resistant effect. 

The dilution-resistant effect has previously been studied in Ref.~\cite{EscuderoAbenza:2020cmq}.
There, it has been shown that a light scalar thermalized after neutrino
decoupling can maximally produce $\Delta N_{{\rm eff}}=0.118$ due
to the dilution-resistant effect. This is below the current experimental
limit but falls in the sensitivity reach of next-generation CMB experiments.
Therefore, once the experimental sensitivity reaches this value, it
will have a great implication: an enormously large part of the parameter
space of light mediators which could be well hidden in the neutrino
sector will be unveiled. 

In this work, we aim at a comprehensive investigation into how future
$N_{{\rm eff}}$ constraints on light mediators in the neutrino sector
might be changed due to the dilution-resistant effect. Our analysis
includes both scalar and vector mediators, and covers a wide mass
range from a few eV to 100 MeV. We concentrate on neutrinophilic light
mediators, but to some extent our results can also be applied to models
like $B-L$. We show that with future experiments such as CMB-S4 and
CMB-HD, $N_{{\rm eff}}$ constraints on such mediators in the sub-MeV
region will be improved by orders of magnitude (see Fig.~\ref{fig:main-result}).
In particular, regarding the recent rising interest in neutrino self-interactions~\cite{Kreisch:2019yzn,Blinov:2019gcj,Berbig:2020wve,Brdar:2020nbj,Deppisch:2020sqh,Das:2022xsz,Bustamante:2020mep,Chang:2022aas,Venzor:2022hql,Venzor:2023aka},
our result implies that strong neutrino self-interactions involving
light mediators can be easily probed or excluded by the next-generation
CMB experiments. 

Our work is structured as follows. Sec.~\ref{sec:Models-and-Boltzmann}
introduces the interactions of the light mediators and the Boltzmann
equations used in this work. The idea of the dilution-resistant effect
is also formulated in this section. In Sec.~\ref{sec:Analytical},
we analytically estimate the cosmological evolution of the light mediators
and provide various formulae that can approximate the numerical results
very well in their respective valid ranges.  Our numerical calculations
and results are presented in Sec.~\ref{sec:result}, where we also
discuss the implications for specific models and neutrino self-interactions.
Finally, we conclude in Sec.~\ref{sec:Conclusions} and relegate
some details to the appendix.

\section{Models and Boltzmann equations\label{sec:Models-and-Boltzmann}}

We consider a light mediator, either a vector denoted by $Z'_{\mu}$
or a scalar denoted by $\phi$, that is coupled to the SM neutrinos
as follows:
\begin{equation}
{\cal L}\supset\begin{cases}
g_{Z'}\nu^{\dagger}\overline{\sigma}^{\mu}\nu Z_{\mu}'\thinspace & \text{for vector}\thinspace,\\
g_{\phi}\nu\nu\phi+{\rm h.c.} & \text{\text{for scalar}}\thinspace.
\end{cases}\label{eq:}
\end{equation}
Throughout this paper, we adopt the notation of two-component Weyl
spinors for neutrinos~\cite{Dreiner:2008tw}.  The masses of $Z_{\mu}'$
and $\phi$ are denoted by $m_{Z'}$ and $m_{\phi}$,  respectively.
For simplicity, we assume that $Z'_{\mu}$ and $\phi$ are coupled
to neutrinos only, and their couplings to charged leptons or quarks
are absent or suppressed. Complete models for such mediators can be
constructed, for example, via the right-handed neutrinos with new
gauge interactions and active-sterile mixing~\cite{Lindner:2013awa,Berbig:2020wve,Chauhan:2020mgv,Chauhan:2022iuh}
or new scalar singlets coupled to right-handed neutrinos~\cite{Xu:2020qek}.
In fact, even for models like $B-L$ in which $Z'$ is equally coupled
to neutrinos and electrons, our analysis below still applies to a
certain extent, as we will discuss later in Sec.~\ref{sec:result}. 

The evolution of a generic species in the expanding universe is governed
by the following Boltzmann equations:
\begin{align}
\frac{dn}{dt}+3Hn & =C_{{\rm prod.}}^{(n)}-C_{{\rm depl.}}^{(n)}\thinspace,\label{eq:-1}\\
\frac{d\rho}{dt}+3H\left(\rho+P\right) & =C_{{\rm prod.}}^{(\rho)}-C_{{\rm depl.}}^{(\rho)}\thinspace,\label{eq:-2}
\end{align}
where $n$, $\rho$, $P$ denote the number, energy, and pressure
densities of the species to be computed\footnote{In our convention, we extract all internal degrees of freedom out
of the definition of $n$ such that $n$ only represents the number
density of a single degree of freedom. For instance, $n_{\nu}=3\zeta(3)T_{\nu}^{3}/(4\pi^{2})$
does not include antineutrinos ($\overline{\nu}$) nor neutrinos of
different flavors. The same convention also applies to $\rho$, $P$,
and the entropy density $s$. }; $H=a^{-1}da/dt$ is the Hubble parameter; and the right-hand sides
are collision terms---see also Appendix~\ref{sec:Collision} for
detailed calculations. The subscripts ``prod.'' and ``depl.''
indicate that the collision terms account for the production and depletion
of the species. 

 Since $dn/dt+3Hn=a^{-3}d\left(na^{3}\right)/dt$ and $H=a^{-1}da/dt$,
we  rewrite Eq.~\eqref{eq:-1} as
\begin{equation}
\frac{d\left(na^{3}\right)}{da}=\frac{a^{2}}{H}\left[C_{{\rm prod.}}^{(n)}-C_{{\rm depl.}}^{(n)}\right].\label{eq:na3}
\end{equation}
For the energy density $\rho$, we obtain a similar equation:
\begin{equation}
\frac{d\left(\rho a^{4}\right)}{da}=\frac{a^{3}}{H}\left[C_{{\rm prod.}}^{(\rho)}-C_{{\rm depl.}}^{(\rho)}\right]+a^{3}\left(\rho-3P\right).\label{eq:rhoa4}
\end{equation}
Compared to Eq.~\eqref{eq:na3}, here we have an extra term proportional
to $\rho-3P$, which vanishes for relativistic species due to the
well-known relation $P=\rho/3$. 

For non-relativistic species, however, this term is always positive,
making a positive contribution to the comoving energy density $\rho a^{4}$
during the Hubble expansion. We refer to it as the dilution-resistant
term, since fundamentally it is exactly this term that causes the
dilution-resistant effect. 

To gain a better understanding of the dilution-resistant term, let
us consider the process $\overline{\nu}\nu\leftrightarrow Z'$. Each
$Z'$ particle being produced via this process consumes one $\nu$
and one $\overline{\nu}$. The collision term $C_{{\rm prod.}}^{(n_{Z'})}$
for $Z'$ production should be exactly equal to the collision terms
for $\nu$ and $\overline{\nu}$ depletion, i.e. $C_{{\rm prod.}}^{(n_{Z'})}=C_{{\rm depl.}}^{(n_{\nu})}=C_{{\rm depl.}}^{(n_{\overline{\nu}})}$.
So when summing Eq.~\eqref{eq:na3} for $Z'$ and $\nu$ together,
we have 
\begin{equation}
\frac{d\left(n_{\nu}a^{3}\right)}{da}+\frac{d\left(n_{Z'}a^{3}\right)}{da}=0\thinspace.\label{eq:-3}
\end{equation}
Following a similar argument, we also obtain
\begin{equation}
\frac{d\left(\rho_{\nu}a^{4}\right)}{da}+\frac{d\left(\rho_{\overline{\nu}}a^{4}\right)}{da}+\frac{d\left(\rho_{Z'}a^{4}\right)}{da}=a^{3}\left(\rho_{Z'}-3P_{Z'}\right).\label{eq:-4}
\end{equation}
Eq.~\eqref{eq:-4} implies that the comoving energy density $(\rho_{\nu}+\rho_{\overline{\nu}}+\rho_{Z'})a^{4}$
would remain  constant if the dilution-resistant term on the right-hand
side were absent.  Due to the presence of this term, we expect that
the total energy of $Z'$ and $\nu$ in a comoving volume should increase
during the Hubble expansion, leading to a positive contribution to
$N_{{\rm eff}}$. As we will show, for a decoupled $Z'$-$\nu$ (or
$\phi$-$\nu$) sector, the dilution-resistant effect can cause $\Delta N_{{\rm eff}}=0.252$
(or $0.118$) maximally.

\section{Analytical estimates\label{sec:Analytical}}

A quantitative and accurate calculation of the dilution-resistant
effect requires numerically solving the Boltzmann equation. Under
certain assumptions, however, most of the numerical results can be
approximately obtained in the analytic approach, as we shall elaborate
below. 

\subsection{Case A: production after neutrino decoupling\label{subsec:Case-A}
}

Let us first consider that the coupling of the mediator is sufficiently
small and its mass is well below the neutrino decoupling temperature.
In this case, the light mediator is in the freeze-in regime and the
production is only significant when the temperature is at the same
order of magnitude of the mass. 

As the light mediator is produced from neutrinos while neutrinos have
decoupled from the thermal bath, there are a few useful conservation
laws which are valid under some circumstances. 
\begin{itemize}
\item Conservation of particle numbers. If $Z'$ is only produced via $\nu\overline{\nu}\to Z'$,
then creating one $Z'$ particle  implies that one $\nu$ and $\overline{\nu}$
must have been destroyed. In a comoving volume, the total number of
$Z'$ and $\nu$  particles should be conserved. For $\nu\nu\leftrightarrow\phi$,
after taking $\overline{\nu}\overline{\nu}\leftrightarrow\phi$ into
account, we obtain a similar conclusion. Therefore, we have the following
conservation law:
\begin{align}
(N_{\nu}n_{\nu}+N_{Z'}n_{Z'})a^{3} & ={\rm constant}\thinspace,\label{eq:-22}\\
(N_{\nu}n_{\nu}+N_{\phi}n_{\phi})a^{3} & ={\rm constant}\thinspace,\label{eq:-23}\\
\text{validity: } & \text{only for }\nu\overline{\nu}\leftrightarrow Z'\text{ and }\nu\nu\leftrightarrow\phi\thinspace,\nonumber 
\end{align}
where
\[
N_{\nu}=3\thinspace,\ N_{Z'}=3\thinspace,\ N_{\phi}=1\thinspace.
\]
Note that Eqs.~\eqref{eq:-22} and \eqref{eq:-23} would be invalid
if $\nu\overline{\nu}\leftrightarrow2Z'$ or $\nu\overline{\nu}\leftrightarrow2\phi$
becomes significant. 
\item Conservation of energy.  If both $Z'$/$\phi$ and $\nu$ are highly
relativistic, the total energy in a comoving volume is conserved:
\begin{align}
(2N_{\nu}\rho_{\nu}+N_{Z'}\rho_{Z'})a^{4} & ={\rm constant}\thinspace,\label{eq:-24}\\
(2N_{\nu}\rho_{\nu}+N_{\phi}\rho_{\phi})a^{4} & ={\rm constant}\thinspace,\label{eq:-25}\\
\text{validity: } & \text{only in the relativistic regime}.\nonumber 
\end{align}
Note that Eqs.~\eqref{eq:-24} and \eqref{eq:-25} would be invalid
if $Z'$ or $\phi$ becomes non-relativistic. 
\item Conservation of entropy. If $Z'$/$\phi$ reaches thermal and chemical
equilibrium with $\nu$, then the total entropy in a comoving volume
is conserved as long as the equilibrium is maintained:
\begin{align}
(2N_{\nu}s_{\nu}+N_{Z'}s_{Z'})a^{3} & ={\rm constant}\thinspace,\label{eq:-26}\\
(2N_{\nu}s_{\nu}+N_{\phi}s_{\phi})a^{3} & ={\rm constant}\thinspace,\label{eq:-27}\\
\text{validity: } & \text{only when \ensuremath{\nu} and \ensuremath{\phi} (\ensuremath{Z'}) are in equilibrium}.\nonumber 
\end{align}
The entropy conservation is only valid when the universe expands slowly
in comparison to particle reaction rates. Equivalently, according
to the second law of thermodynamics, the process has to be reversible
(i.e. if the universe shrinks back, the same thermodynamic status
can be recovered) to guarantee that the comoving entropy does not
increase. Therefore, for freeze-in processes where the universe expands
faster than particle reaction rates, the entropy conservation is not
applicable. 
\end{itemize}

\subsubsection{The evolution}

For simplicity, our analysis below will be concentrated on the vector
case. The generalization to the scalar case is straightforward and
the corresponding analytic results will also be presented. 

Let us first consider that the coupling $g_{Z'}$ is sufficiently
small and $m_{Z'}$ is well below the neutrino decoupling temperature.
In this case, $Z'$ is in the freeze-in regime and the production
is significant only when the temperature is at the same order of magnitude
of $m_{Z'}$. 

Since $g_{Z'}$ is small, the dominant process for $Z'$ production
is $\nu\overline{\nu}\to Z'$. Other processes like $\nu\overline{\nu}\to2Z'$
are suppressed by higher orders of $g_{Z'}$. In the Boltzmann approximation,
the collision term for $Z'$ production is given by
\begin{equation}
C_{{\rm prod.}}^{(n_{Z'})}=\sum_{\alpha}C_{\nu_{\alpha}\overline{\nu_{\alpha}}\to Z'}^{(n_{Z'})}=N_{\nu}\frac{|{\cal M}|^{2}}{32\pi^{3}}m_{Z'}T_{\nu}K_{1}\left(\frac{m_{Z'}}{T_{\nu}}\right),\label{eq:-5}
\end{equation}
where $\alpha$ denotes neutrino flavors, $T_{\nu}$ is the neutrino
temperature, and $|{\cal M}|^{2}$ represents the squared matrix element
of $\nu\overline{\nu}\leftrightarrow Z'$:
\begin{equation}
|{\cal M}|^{2}=\frac{2}{3}g_{Z'}^{2}m_{Z'}^{2}\thinspace.\label{eq:-6}
\end{equation}

At $T_{\nu}\gg m_{Z'}$, the back-reaction $\nu\overline{\nu}\leftarrow Z'$
is negligible due to $n_{\nu}\gg n_{Z'}$. In this regime, the number
density can be estimated by directly integrating over $C_{{\rm prod.}}^{(n_{Z'})}$:
\begin{align}
n_{Z'} & \approx T_{\nu}^{3}\frac{m_{{\rm pl}}}{g_{H\nu}}\int_{T_{\nu}}^{\infty}C_{{\rm prod.}}^{(n_{Z'})}\tilde{T}_{\nu}^{-6}d\tilde{T_{\nu}}\label{eq:-7}\\
 & \approx N_{\nu}\frac{|{\cal M}|^{2}m_{\text{pl}}}{96\pi^{3}g_{H\nu}}\thinspace,\label{eq:-7-1}
\end{align}
where $g_{H\nu}\equiv m_{{\rm pl}}H/T_{\nu}^{2}$ and $m_{{\rm pl}}=1.22\times10^{19}$
GeV is the Planck mass. After neutrino decoupling, $g_{H\nu}$ is
approximately a constant, $g_{H\nu}\approx6$.  Eq.~\eqref{eq:-7}
takes a freeze-in formula from \cite{Li:2022bpp}. From Eq.~\eqref{eq:-7}
to Eq.~\eqref{eq:-7-1}, we have used $K_{1}\left(m_{Z'}/T_{\nu}\right)\approx T_{\nu}/m_{Z'}+{\cal O}(m_{Z'}/T_{\nu})$
in the $T_{\nu}\gg m_{Z'}$ regime.

\begin{figure}
\centering

\includegraphics[width=0.8\textwidth]{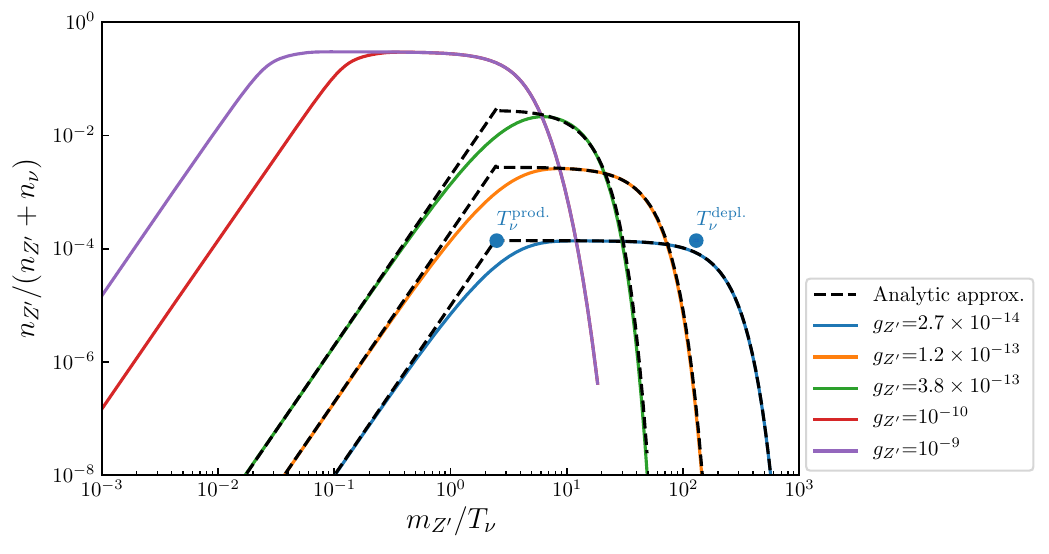}

\caption{The evolution of $n_{Z'}/(n_{Z'}+n_{\nu})$ obtained via numerical
calculations, compared with analytic approximations (dashed lines)
obtained from Eq.~\eqref{eq:-17}. The two blues points marked by
$T_{\nu}^{\text{prod.}}$ and $T_{\nu}^{\text{depl.}}$ represent
the end of $Z'$ production and the beginning of $Z'$ depletion,
computed using Eqs.~\eqref{eq:-11} and \eqref{eq:-10}. The mass
of $Z'$ in the shown examples is $m_{Z'}=1$ keV. \label{fig:approx}}
\end{figure}

Despite that the result in Eq.~\eqref{eq:-7} appears as a temperature-independent
constant, the comoving number density $n_{Z'}a^{3}$ actually keeps
increasing as the universe expands. The comoving number density stops
increasing when the neutrino temperature is insufficient to produce
$Z'$. The maximum of $n_{Z'}a^{3}$ in the small $g_{Z'}$ limit
can be obtained by replacing $\int_{T_{\nu}}^{\infty}\to\int_{0}^{\infty}$
in Eq.~\eqref{eq:-7}.  This corresponds to the assumption that the
decay of $Z'$ starts only after the freeze-in production completes.
The result is
\begin{equation}
n_{Z'}^{{\rm max}}\approx\frac{3N_{\nu}|{\cal M}|^{2}T_{\nu}^{3}m_{\text{pl}}}{64\pi^{2}g_{H\nu}m_{Z'}^{3}}\thinspace.\label{eq:-8}
\end{equation}
By equating Eq.~\eqref{eq:-8} to Eq.~\eqref{eq:-7}, we obtain the
following  temperature
\begin{equation}
T_{\nu}^{{\rm prod.}}\approx\left(\frac{2}{9\pi}\right)^{1/3}m_{Z'}\approx0.4m_{Z'}\thinspace,\label{eq:-11}
\end{equation}
which can be roughly taken as the temperature when the production
completes---see the blue point marked by ``$T_{\nu}^{{\rm prod.}}$''
in Fig.~\ref{fig:approx}. 

After  the production completes, $n_{Z'}a^{3}$ will remain constant
for a while until the depletion term in Eq.~\eqref{eq:na3} becomes
significant. For the depletion term to be significant, at least the
age of the universe $\tau_{\text{universe}}\sim1/H$ needs to be longer
than the lifetime of $Z'$ at rest, $\tau_{Z'}=1/\Gamma_{Z'}$ where
\begin{equation}
\Gamma_{Z'}=N_{\nu}\frac{|{\cal M}|^{2}}{16\pi m_{Z'}}\thinspace.\label{eq:-9}
\end{equation}
Therefore, $\tau_{\text{universe}}\gtrsim\tau_{Z'}$ implies that
$T_{\nu}$ needs to  be below
\begin{equation}
T_{\nu}^{\text{depl.}}\approx\left(\frac{N_{\nu}|{\cal M}|^{2}m_{{\rm pl}}}{16\pi g_{H\nu}m_{Z'}}\right)^{1/2},\label{eq:-10}
\end{equation}
which is obtained by solving $H\approx\Gamma_{Z'}$. Eq.~\eqref{eq:-10}
can be taken as the temperature when the depletion begins---see the
blue point marked by ``$T_{\nu}^{{\rm depl.}}$'' in Fig.~\ref{fig:approx}.
After that, the comoving number density decays exponentially. Its
evolution can be computed by solving Eq.~\eqref{eq:na3} with the
production term neglected, i.e., $d\left(n_{Z'}a^{3}\right)/da=-a^{2}C_{{\rm depl.}}^{(n_{Z'})}/H$,
 where $C_{{\rm depl.}}^{(n_{Z'})}$ takes the non-relativistic approximation
(see Appendix~\ref{sec:Collision}): 
\begin{equation}
C_{{\rm depl.}}^{(n_{Z'})}\approx\Gamma_{Z'}n_{Z'}\thinspace.\label{eq:-13}
\end{equation}
By defining $X\equiv n_{Z'}a^{3}$, we can rewrite Eq.~\eqref{eq:na3}
as 
\begin{equation}
\frac{dX}{da}=-\xi aX,\ \ \xi\equiv\frac{\Gamma_{Z'}m_{\text{pl}}}{T_{a}^{2}g_{H\nu}}\thinspace,\label{eq:-12}
\end{equation}
where $X\equiv n_{Z'}a^{3}$ and $T_{a}\equiv T_{\nu}a$. Note that
$T_{a}$ is a constant because $T_{\nu}$ scales as $a^{-1}$. The
above differential equation has the following simple solution:
\begin{equation}
X\propto e^{-\frac{\xi}{2}a^{2}}\thinspace.\label{eq:-14}
\end{equation}
The initial value is determined by Eq.~\eqref{eq:-8}. 

Assembling the above pieces, we obtain the following analytic result
for the evolution of $n_{Z'}$: 
\begin{equation}
n_{Z'}=n_{Z'}^{\max}\times\begin{cases}
\frac{2}{9\pi T_{\nu}^{3}}m_{Z'}^{3} & \text{for }T_{\nu}>T_{\nu}^{{\rm prod.}}\\[2mm]
\exp\left[-\frac{T_{\nu}^{-2}-\left(T_{\nu}^{\text{prod.}}\right)^{-2}}{2g_{H\nu}}\Gamma_{Z'}m_{\text{pl}}\right] & \text{for }T_{\nu}\leq T_{\nu}^{\text{prod.}}
\end{cases}\ .\label{eq:-17}
\end{equation}

Fig.~\ref{fig:approx} shows how well the analytic result in Eq.~\eqref{eq:-17}
approximates the actual evolution of $n_{Z'}$ obtained from numerical
calculations. 

For the scalar case, following similar steps, we obtain
\begin{equation}
n_{\phi}=n_{\phi}^{\max}\times\begin{cases}
\frac{2}{9\pi T_{\nu}^{3}}m_{\phi}^{3} & \text{for }T_{\nu}>T_{\nu}^{{\rm prod.}}\\[2mm]
\exp\left[-\frac{T_{\nu}^{-2}-\left(T_{\nu}^{\text{prod.}}\right)^{-2}}{2g_{H\nu}}\Gamma_{\phi}m_{\text{pl}}\right] & \text{for }T_{\nu}\leq T_{\nu}^{\text{prod.}}
\end{cases}\ ,\label{eq:-63}
\end{equation}
where $n_{\phi}^{\max}$, $\Gamma_{\phi}$, and $T_{\nu}^{{\rm prod.}}$
are almost the same as Eqs.~\eqref{eq:-8}, \eqref{eq:-9} and \eqref{eq:-11}
except that $m_{Z'}$ should be replaced by $m_{\phi}$. The main
difference is in the squared amplitude, which for the scalar should
be 
\begin{equation}
|{\cal M}|^{2}=g_{\phi}^{2}m_{\phi}^{2}\thinspace.\label{eq:-64}
\end{equation}
The difference between $N_{\phi}=1$ and $N_{Z'}=3$ is not of concern
here because in our convention $n_{\phi}$ and $n_{Z'}$ are only
for single degree of freedom.

\subsubsection{The dilution-resistant effect and $\Delta N_{{\rm eff}}$}

Having obtained the cosmological evolution of $Z'$, we then employ
Eq.~\eqref{eq:-4} to estimate the dilution-resistant effect:
\begin{equation}
\frac{d\left(\rho_{\text{inv}}a^{4}\right)}{da}=N_{Z'}a^{3}\left(\rho_{Z'}-3P_{Z'}\right)\thinspace,\label{eq:-16}
\end{equation}
where $\rho_{\text{inv}}$ denotes the total energy density of the
invisible sector ($Z'+\nu$), including neutrinos with three flavors
and $Z'$ with three polarizations.  Note that $N_{{\rm eff}}$ as
a CMB observable is defined as
\begin{equation}
N_{{\rm eff}}\equiv\frac{\rho_{{\rm inv}}}{2\rho_{\nu}^{{\rm st}}}\thinspace,\label{eq:Neff-def}
\end{equation}
where $\rho_{\nu}^{{\rm st}}$ denotes the neutrino energy density
in the SM of a single flavor, and the factor of $2$ in front of it
comes from combining neutrinos and antineutrinos. According to Eqs.~\eqref{eq:-16}
and \eqref{eq:Neff-def}, the contribution of the dilution-resistant
effect to $N_{{\rm eff}}$  can be computed as follows: 
\begin{equation}
\Delta N_{{\rm eff}}=\frac{\Delta}{2\rho_{\nu}^{{\rm st}}a^{4}},\ \ \Delta\equiv N_{Z'}\int_{a_{0}}^{a_{1}}a^{3}\left(\rho_{Z'}-3P_{Z'}\right)da\thinspace.\label{eq:-20}
\end{equation}
Here $a_{0}$ and $a_{1}$ denote the values of $a$ at two generic
moments, and Eq.~\eqref{eq:-20} only accounts for the contribution
of the period when the universe expands from $a=a_{0}$ to $a=a_{1}$.
In practice, to compute the contribution of the entire relevant period,
we can set $a_{0}$ at a moment when $Z'$ has not been significantly
produced, and $a_{1}$ at the moment of the recombination. 

Here we only consider  the period when $T_{\nu}\leq T_{\nu}^{{\rm prod.}}$
so only the exponential decay part of Eq.~\eqref{eq:-17} will be
used. According to Eq.~\eqref{eq:-11}, we take the non-relativistic
approximation ($\rho_{Z'}\approx n_{Z'}m_{Z'}$, $P_{Z'}\approx0$)
and obtain
\begin{equation}
\Delta\approx\int_{a_{0}}^{\infty}3a^{3}n_{Z'}m_{Z'}da\approx\frac{9T_{a}^{4}}{4\sqrt{2\pi}m_{Z'}}e^{\frac{\xi_{0}}{2}}\text{erfc}\left(\sqrt{\frac{\xi_{0}}{2}}\right)T_{\nu}^{{\rm prod.}}\xi_{0}^{1/2}\thinspace,\label{eq:-18}
\end{equation}
where $\xi_{0}\equiv\xi a_{0}^{2}$ should be a small number ($\xi_{0}\ll1$)
if $g_{Z'}$ is sufficiently small. So  we expand Eq.~\eqref{eq:-18}
in terms of $\xi_{0}$ and take the leading order:  
\begin{equation}
\Delta\approx\frac{9T_{a}^{4}}{4\sqrt{2\pi}m_{Z'}}T_{\nu}^{{\rm prod.}}\xi_{0}^{1/2}\approx\frac{9T_{a}^{4}}{4m_{Z'}}\sqrt{\frac{\Gamma_{Z'}m_{{\rm pl}}}{2\pi g_{H\nu}}}\thinspace.\label{eq:-19}
\end{equation}

Substituting it into Eq.~\eqref{eq:-20}, we obtain
\begin{equation}
\Delta N_{{\rm eff}}\approx\frac{9T_{\nu}^{4}}{8m_{Z'}\rho_{\nu}^{{\rm st}}}\sqrt{\frac{\Gamma_{Z'}m_{{\rm pl}}}{2\pi g_{H\nu}}}\approx0.04\cdot\left(\frac{g_{Z'}}{10^{-13}}\right)\cdot\left(\frac{m_{Z'}}{\text{keV}}\right)^{-1/2}.\label{eq:-21}
\end{equation}

\begin{figure}
\centering

\includegraphics[width=0.49\textwidth]{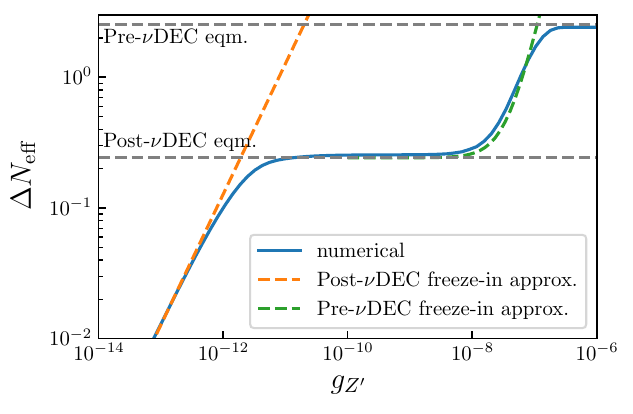}\includegraphics[width=0.49\textwidth]{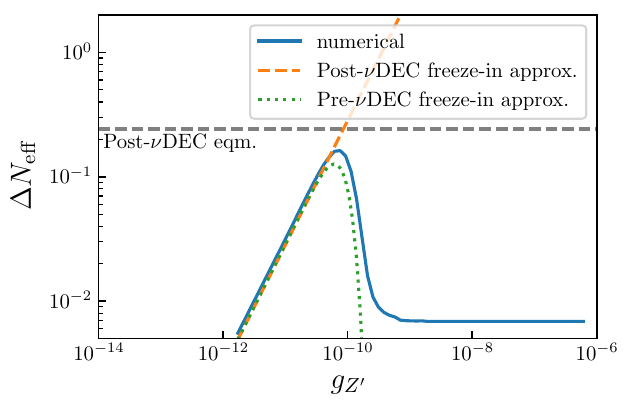}

\caption{$\Delta N_{{\rm eff}}$ as a function of $g_{Z'}$. Left panel: $m_{Z'}=10$
keV. Right panel: $m_{Z'}=20$ MeV. The dashed curves labeled ``Post-$\nu$DEC
freeze-in approx.'' represent the analytic approximate result in
Eq.~\eqref{eq:-21}. The ``Pre-$\nu$DEC freeze-in approx.'' curves
take the approximate expressions in Eq.~\eqref{eq:small-m-pre} for
the left panel and Eq.~\eqref{eq:large-m-pre} for the right.  \label{fig:approx-Neff}}
\end{figure}

Fig.~\ref{fig:approx-Neff} compares Eq.~\eqref{eq:-21} with the
actual result obtained from numerical calculations. As is expected,
the approximate formula agrees well the numerical result in the small
$g_{Z'}$ limit. In this limit, $\Delta N_{{\rm eff}}$ increases
linearly with $g_{Z'}$. When $g_{Z'}$ increases to a certain value,
$Z'$ will reach equilibrium with $\nu$, and the above calculation
is no longer applicable. The calculation dealing with equilibrium
will be presented in Sec~\ref{subsec:Maximal}.

For the scalar case, $\Delta$ is reduced by a factor of three due
to $N_{Z'}\to N_{\phi}$, and $|{\cal M}|^{2}$ should take the expression
in Eq.~\eqref{eq:-64}. The final result for the scalar case is
\begin{equation}
\Delta N_{{\rm eff}}\approx\frac{3T_{\nu}^{4}}{8m_{\phi}\rho_{\nu}^{{\rm st}}}\sqrt{\frac{\Gamma_{\phi}m_{{\rm pl}}}{2\pi g_{H\nu}}}\approx0.017\cdot\left(\frac{g_{\phi}}{10^{-13}}\right)\cdot\left(\frac{m_{\phi}}{\text{keV}}\right)^{-1/2}.\label{eq:-21-1}
\end{equation}

\subsubsection{The equilibrium values of $\Delta N_{{\rm eff}}$\label{subsec:Maximal}}

\begin{table*}
\centering 

\begin{tabular}{ccccc}
\toprule 
 &  & scalar & vector & \tabularnewline
\midrule 
 & Post-$\nu$DEC equilibrium  & $\Delta N_{{\rm eff}}=0.118$ & $\Delta N_{{\rm eff}}=0.242$ & \tabularnewline
 & Pre-$\nu$DEC equilibrium & $\Delta N_{{\rm eff}}=0.794$ & $\Delta N_{{\rm eff}}=2.53$ & \tabularnewline
 & Pre-$\nu$DEC equilibrium (strong couplings) & $\Delta N_{{\rm eff}}=0.785$ & $\Delta N_{{\rm eff}}=2.48$ & \tabularnewline
\bottomrule
\end{tabular}


\caption{\label{tab:t} Equilibrium values of $\Delta N_{{\rm eff}}$  assuming
$Z'$ or $\phi$ reaches equilibrium with neutrinos before (Pre-$\nu$DEC)
or after (Post-$\nu$DEC) neutrino decoupling. The last row applies
to strong couplings which can maintain equilibrium via $\nu\overline{\nu}\leftrightarrow2Z'$
or $\nu\overline{\nu}\leftrightarrow2\phi$. Compared to the second
row where only $\nu\overline{\nu}\leftrightarrow Z'$ or $\nu\nu\leftrightarrow\phi$
are in equilibrium so that Eqs.~\eqref{eq:-22} and \eqref{eq:-23}
are satisfied, the last row only requires entropy conservation.  
The values presented in this table are only applicable to the mass
range $1\ {\rm eV}\ll m_{Z'/\phi}\ll1\ \text{MeV}$. }
\end{table*}

As already shown in Fig.~\ref{fig:approx-Neff}, when Eq.~\eqref{eq:-21}
becomes invalid at large $g_{Z'}$, $\Delta N_{{\rm eff}}$ stays
at a constant which is about $0.242$. This is actually the maximal
value the dilution-resistant effect could cause if one requires that
$Z'$ is only produced from decoupled neutrinos. If $g_{Z'}$ further
increases, $Z'$ can thermalize before neutrino decoupling so that
$\Delta N_{{\rm eff}}$ will exceed this value and increase to another
constant level at $2.53$. We refer to these numbers as the equilibrium
values of $\Delta N_{{\rm eff}}$.

The equilibrium values of $\Delta N_{{\rm eff}}$ in their respective
valid ranges are almost independent of the coupling and the mass,
and can be computed simply from equilibrium conditions and the conservation
laws introduced at the beginning of Sec.~\ref{subsec:Case-A}. Below
we present the calculation.

For $Z'$ reaching equilibrium with $\nu$ after neutrino decoupling,
which we refer to as Post-$\nu$DEC equilibrium, the conservation
laws in Eqs.~\eqref{eq:-22}, \eqref{eq:-24}, and \eqref{eq:-26}
can be written as
\begin{align}
(N_{\nu}n_{\nu1}+0)a_{1}^{3} & =(N_{\nu}n_{\nu2}+N_{Z'}n_{Z'2})a_{2}^{3}=(N_{\nu}n_{\nu3}+0)a_{3}^{3}\thinspace,\label{eq:-28}\\
(2N_{\nu}\rho_{\nu1}+0)a_{1}^{4} & =(2N_{\nu}\rho_{\nu2}+N_{Z'}\rho_{Z'2})a_{2}^{4}\thinspace,\label{eq:-29}\\
(2N_{\nu}s_{\nu2}+N_{Z'}s_{Z'2})a_{2}^{3} & =(2N_{\nu}s_{\nu3}+0)a_{3}^{3}\thinspace,\label{eq:-30}
\end{align}
where the subscripts ``1, 2, 3'' denote  three phases when (1) $Z'$
has not been significantly produced; (2) $Z'$ reaches equilibrium
and keeps relativistic; (3) $Z'$ has completely decayed.  

From Eqs.~\eqref{eq:-28} and \eqref{eq:-29}, we have
\begin{align}
n_{\text{FD}}(T_{1},0)a_{1}^{3} & =\left[n_{\text{FD}}(T_{2},\mu_{2})+n_{\text{BE}}(T_{2},2\mu_{2})\right]a_{2}^{3}\thinspace,\label{eq:-31}\\
2\rho_{\text{FD}}(T_{1},0)a_{1}^{4} & =\left[2\rho_{\text{FD}}(T_{2},\mu_{2})+\rho_{\text{BE}}(T_{2},2\mu_{2})\right]a_{2}^{4}\thinspace,\label{eq:-32}
\end{align}
where $n_{\text{FD/BE}}$ and $\rho_{\text{FD/BE}}$ denote the number
and energy densities of massless particles in Fermi-Dirac/Bose-Einstein
distributions, given as follows
\begin{align}
n_{\text{FD/BE}}(T,\mu) & \equiv\int\frac{1}{e^{(p-\mu)/T}\pm1}\frac{d^{3}p}{\left(2\pi\right)^{3}}=\mp\frac{T^{3}}{\pi^{2}}\text{Li}_{3}(\mp e^{\mu/T}),\label{eq:-33}\\
\rho_{\text{FD/BE}}(T,\mu) & \equiv\int\frac{p}{e^{(p-\mu)/T}\pm1}\frac{d^{3}p}{\left(2\pi\right)^{3}}=\mp\frac{3T^{4}}{\pi^{2}}\text{Li}_{4}(\mp e^{\mu/T}).\label{eq:-34}
\end{align}
Here $\text{Li}_{3,4}$ are polylogarithm functions. 

Substituting Eqs.~\eqref{eq:-33} and \eqref{eq:-34} into Eqs.~\eqref{eq:-31}
and \eqref{eq:-32}, and solving the equations, we obtain
\begin{equation}
\left(T_{2},\ \mu_{2}\right)=\left(1.208,-1.166\right)T_{1}\frac{a_{1}}{a_{2}}\thinspace.\label{eq:-35}
\end{equation}

Similarly, we can solve the equations that connect the second phase
to the third phase. The entropy density is computed by\footnote{Eq.~\eqref{eq:-36} only applies to equilibrium distributions. For
non-equilibrium distributions, we refer to Ref.~\cite{Grohs:2015tfy}
for a more general definition, which reduces to Eq.~\eqref{eq:-36}
in the equilibrium case. }
\begin{equation}
s=\frac{\rho+P-\mu n}{T}=\frac{4\rho/3-\mu n}{T}\thinspace.\label{eq:-36}
\end{equation}
The result is
\begin{equation}
\left(T_{3},\ \mu_{3}\right)=\left(1.092,-0.3133\right)T_{1}\frac{a_{1}}{a_{3}}\thinspace.\label{eq:-37}
\end{equation}
Using Eq.~\eqref{eq:-37} to compute the final energy density of neutrinos,
$\rho_{\nu3}$, we obtain 
\begin{equation}
\Delta N_{\text{eff}}=3\left[\frac{\rho_{\nu3}a_{3}^{4}}{\rho_{\nu1}a_{1}^{4}}-1\right]=0.242\thinspace.\label{eq:-38}
\end{equation}

For the scalar case, the calculation is similar except that we need
to replace $N_{Z'}=3\to N_{\phi}=1$. This leads to $\Delta N_{\text{eff}}=0.118$,
which reproduces the previous result obtained in Ref.~\cite{EscuderoAbenza:2020cmq}. 

For $Z'$ reaching equilibrium before neutrino decoupling, which we
refer to as Pre-$\nu$DEC equilibrium, the analysis is simpler---we
only need to solve two equations (one for $n$ and the other for $s$)
connecting the second and the third phases. During the second phase,
the chemical potential $\mu_{2}$ remains zero because all particles
are in equilibrium with photons and the reaction rate of $\gamma+e^{\pm}\to N\gamma+e^{\pm}$
is high. By solving the equations for $n$ and $s$, we obtain
\begin{equation}
\left(T_{3},\ \mu_{3}\right)=\left(1.326,-0.715\right)T_{2}\frac{a_{2}}{a_{3}}\thinspace,\label{eq:-39}
\end{equation}
which gives $\Delta N_{\text{eff}}=2.53$. 

If the coupling $g_{Z'}$ is sufficiently strong so that both $\nu\overline{\nu}\leftrightarrow Z'$
and $\nu\overline{\nu}\leftrightarrow2Z'$ are in equilibrium, then
the conservation of particle numbers is violated. In this case, we
only need to solve the equation for $s$, with zero chemical potentials
because the two reactions in equilibrium imply $\mu_{\nu}+\mu_{\overline{\nu}}=\mu_{Z'}$
and $\mu_{\nu}+\mu_{\overline{\nu}}=2\mu_{Z'}$. The result for this
case is slightly different, $\Delta N_{\text{eff}}=2.48$.

In Tab.~\ref{tab:t}, we summarize all equilibrium values of $\Delta N_{\text{eff}}$
for the aforementioned cases, including both scalar and vector cases.

\subsection{Case B: production before neutrino decoupling }

For large $g_{Z'}$, or large $m_{Z'}$, the production of $Z'$ before
neutrino decoupling, i.e. Pre-$\nu$DEC production, is important.
The calculations, and hence the results, are very different for $m_{Z'}\lesssim T_{\nu}^{{\rm dec}}$
and $m_{Z'}\gtrsim T_{\nu}^{{\rm dec}}$ where $T_{\nu}^{{\rm dec}}$
is the neutrino decoupling temperature.

Let us first consider $m_{Z'}\ll T_{\nu}^{{\rm dec}}$. Although the
production at temperatures above $T_{\nu}^{{\rm dec}}$ is suppressed
by $C_{{\rm prod.}}^{(n_{Z'})}/T_{\nu}^{4}\propto m_{Z'}^{2}/T_{\nu}^{2}$,
one still gets a small amount of $Z'$ particles produced before neutrino
decoupling. The energy density of $Z'$ being produced before neutrino
decoupling can be computed by integrating $C_{{\rm prod.}}^{(\rho_{Z'})}$
as follows {[}similar to Eq.~\eqref{eq:-7} for $n_{Z'}${]}: 
\begin{equation}
\rho_{Z'}=T_{\nu}^{4}\frac{m_{{\rm pl}}}{g_{H\nu}}\int_{T_{\nu}}^{\infty}C_{{\rm prod.}}^{(\rho_{Z'})}\tilde{T}_{\nu}^{-7}d\tilde{T_{\nu}}=\left.N_{\nu}T_{\nu}\frac{|{\cal M}|^{2}m_{{\rm pl}}}{48\pi^{3}g_{H\nu}}\right|_{T_{\nu}\to T_{\nu}^{{\rm dec}}}\thinspace.\label{eq:-40}
\end{equation}
Here $g_{H\nu}$ is slightly different from that in the Post-$\nu$DEC
epoch, $g_{H\nu}=5.44$. The energy in Eq.~\eqref{eq:-40}, which
does not cost any Post-$\nu$DEC neutrinos, will eventually be injected
into the decoupled neutrino sector. So the contribution to $N_{{\rm eff}}$
is 
\begin{equation}
\Delta N_{{\rm eff}}=\Delta N_{{\rm eff}}^{{\rm DR}}+\left.\frac{N_{Z'}\rho_{Z'}}{2\rho_{\nu}^{{\rm st.}}}\right|_{T_{\nu}\to T_{\nu}^{{\rm dec}}}\approx0.242+1.86\times\left(\frac{g}{10^{-7}}\frac{m_{Z'}}{10\text{ keV}}\right)^{2}\thinspace,\label{eq:small-m-pre}
\end{equation}
where $\Delta N_{{\rm eff}}^{{\rm DR}}\approx0.242$ is caused by
the dilution-resistant effect derived in Sec.~\ref{subsec:Maximal}.
Eq.~\eqref{eq:small-m-pre} is plotted in the left panel of Fig.~\ref{fig:approx-Neff},
shown as the green dashed curve, which approximates well the second
rise of the numerical curve. 

Next, we turn to the case of $m_{Z'}\gg T_{\nu}^{{\rm dec}}$. If
the coupling is strong, then $Z'$ would be in thermal equilibrium
and its abundance would be Boltzmann suppressed at neutrino decoupling,
i.e. $n_{Z'}\propto\exp(-m_{Z'}/T_{\nu}^{{\rm dec}})$. If the coupling
is very weak, then the abundance of $Z'$ at neutrino decoupling is
also suppressed due to insufficient production. Therefore, the dependence
of $\Delta N_{{\rm eff}}$ on $g_{Z'}$ is not monotonic, as one can
see in the right panel of Fig.~\ref{fig:approx-Neff}. Nevertheless,
we would like to point out here that the Post-$\nu$DEC freeze-in
formula in Eq.~\eqref{eq:-21}, i.e. the orange dashed curve, can
still fit the low-$g_{Z'}$ part of the blue curve very well. This
is because in this regime, $Z'$ is long-lived and most of the produced
$Z'$ only decay after neutrino decoupling. So in the small $g_{Z'}$
limit, the dominant contribution is still from the dilution-resistant
effect. 

When $g_{Z'}$ is large, we need to take into account both the dilution-resistant
effect, which only starts after neutrino decoupling, and the amount
of energy that the invisible sector has gained from the SM thermal
bath before neutrino decoupling. The former can be obtained by repeating
the calculation in Sec.~\ref{subsec:Case-A} with some minor changes:
(i) the integration should start from $T_{\nu}=T_{\nu}^{{\rm dec}}\approx2\ \text{MeV}$
instead of $T_{\nu}=T_{\nu}^{{\rm prod}}$; (ii) $g_{H\nu}$ is changed
to $5.44$; (iii) in Eq.~\eqref{eq:-17} we take $T_{\nu}^{-2}-(T_{\nu}^{\text{prod.}})^{-2}\approx T_{\nu}^{-2}$.
The latter can be computed using Eq.~\eqref{eq:-17} and the non-relativistic
approximation $\rho_{Z'}\approx n_{Z'}m_{Z'}$ at $T_{\nu}=T_{\nu}^{{\rm dec}}$.
Combining the two contributions, we obtain

\begin{equation}
\Delta N_{{\rm eff}}\approx0.4\cdot g_{m}\left[1-\text{erf}(g_{m})\right]\left(\frac{10\text{MeV}}{m_{Z'}}\right)+0.5\cdot g_{m}^{2}e^{-g_{m}^{2}}\left(\frac{10\text{MeV}}{m_{Z'}}\right)\thinspace,\label{eq:large-m-pre}
\end{equation}
where 
\begin{equation}
g_{m}\equiv\left(\frac{g_{Z'}}{10^{-10}}\right)\left(\frac{m_{Z'}}{10\ {\rm MeV}}\right)^{1/2}.\label{eq:-41}
\end{equation}
For $g_{m}\ll1$, we have $1-\text{erf}(g_{m})\approx1-2g_{m}/\sqrt{\pi}$
so $\Delta N_{{\rm eff}}$ in Eq.~\eqref{eq:large-m-pre} is dominated
by its first term, which corresponds to the dilution-resistant effect.
As is shown in the right panel of Fig.~\ref{fig:approx-Neff}, in
the small $g_{Z'}$ limit, $\Delta N_{{\rm eff}}$ increases linearly
with $g_{Z'}$. For $g_{m}\gg1$, we have $1-\text{erf}(g_{m})\approx e^{-g_{m}^{2}}/(\sqrt{\pi}g_{m})$
so $\Delta N_{{\rm eff}}$ in Eq.~\eqref{eq:large-m-pre} is suppressed
by $e^{-g_{m}^{2}}$. The linear increase combined with the exponential
decrease explains the non-monotonic behavior of the $\Delta N_{{\rm eff}}$-$g_{Z'}$
curve.

\section{Numerical calculations and results\label{sec:result}}

The Boltzmann equations of neutrinos and $Z'$ ($\phi$) can be solved
numerically, though some technical issues such as the stiffness of
differential equations and the overflow of floating-point numbers
encountered in overlarge Boltzmann suppression might affect the stability
of the numerical solutions. These technical issues are discussed in
Appendix~\ref{sec:Technical}. 

Another important issue is neutrino decoupling which splits the SM
plasma into two decoupled sectors. To include neutrino decoupling
in our numerical calculation without solving additional Boltzmann
equations, we assume instant neutrino decoupling. Before neutrino
decoupling, we use $T_{\nu}=T_{\gamma}\propto a^{-1}$ to determine
the neutrino temperature and the Hubble expansion rate. So essentially
we only need to solve the Boltzmann equation for $\phi$/$Z'$. After
neutrino decoupling, we solve the Boltzmann equations for both neutrinos
and $\phi$/$Z'$, while the photon-electron sector is calculated
using entropy conservation. 

With appropriate treatments of the aforementioned issues, it is straightforward
to solve the Boltzmann equations numerically\footnote{Our code is publicly available at \url{https://github.com/xunjiexu/Neff-light-Z}.}. 

 For each given sample of $(g_{Z'},\ m_{Z'})$ or $(g_{\phi},\ m_{\phi})$,
we set the beginning of the numerical solution at a sufficiently high
temperature which should be not only well above the mediator mass,
but also above $T_{\nu}^{{\rm dec}}$. Then we evolve the $\nu$-$Z'$
or $\nu$-$\phi$ coupled system according to Eqs.~\eqref{eq:na3}
and \eqref{eq:rhoa4} down to a sufficiently low temperature. Before
neutrino decoupling, the abundance of neutrinos is not affected by
$\nu$-$Z'$ or $\nu$-$\phi$ reactions. This is implemented in our
code by switching off the $\nu$-$Z'$ ($\phi$ ) reactions on the
right-hand sides of Eqs.~\eqref{eq:na3} and \eqref{eq:rhoa4} for
neutrinos if $T_{\nu}>T_{\nu}^{{\rm dec}}$. In doing so, neutrinos
are not affected by the presence of $Z'$ or $\phi$ before decoupling,
while the latter is affected by the former. 

\begin{figure}
\centering

\includegraphics[width=0.95\textwidth]{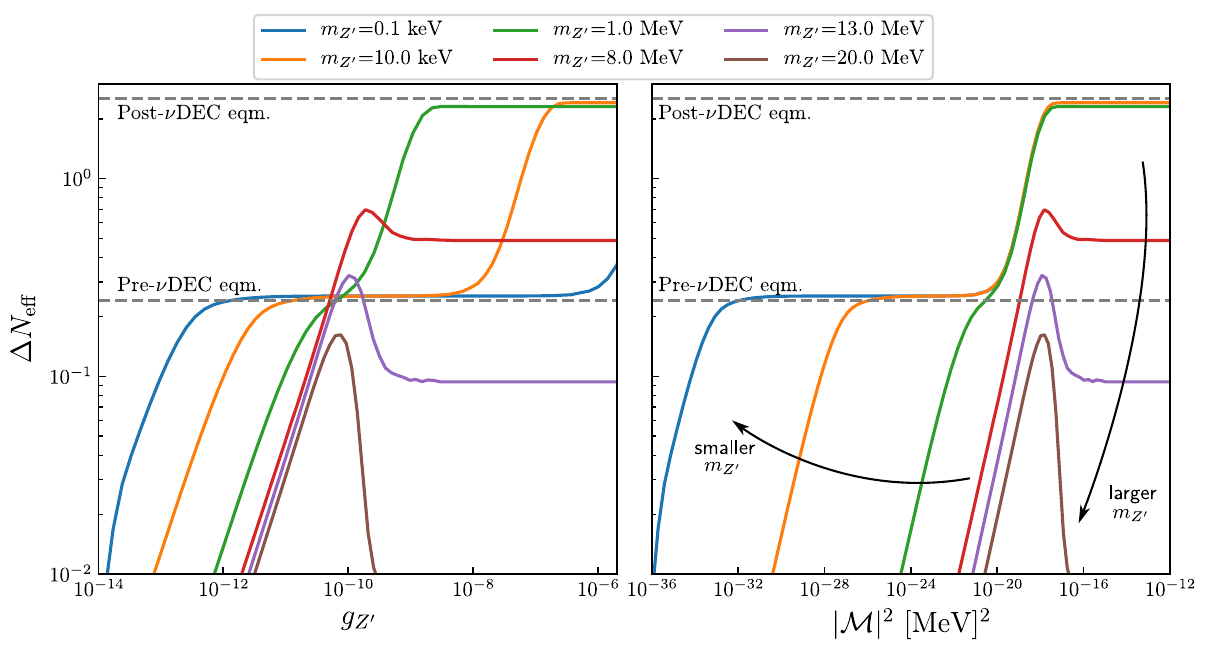}

\caption{$\Delta N_{{\rm eff}}$ computed by solving the Boltzmann equations
numerically for several examples. The left and right panels present
$\Delta N_{{\rm eff}}$ as functions of $g_{Z'}$ and $|{\cal M}|^{2}=2g_{Z'}^{2}m_{Z'}^{2}/3$,
respectively. \label{fig:Neff-examples}}
\end{figure}

Fig.~\ref{fig:Neff-examples} shows the results of $\Delta N_{{\rm eff}}$
obtained from our numerical calculation for several selected values
of $m_{Z'}$. As is expected from previous discussions in Sec.~\ref{subsec:Maximal},
some curves become flat (i.e.~$g_{Z'}$-independent) in certain ranges
of $g_{Z'}$ and the corresponding values $\Delta N_{{\rm eff}}$
(i.e. the gray dashed lines) are approximately given by the equilibrium
values in Tab.~\ref{tab:t}. For those examples that can reach the
two equilibrium values of $\Delta N_{{\rm eff}}$ indicated by the
two gray dashed lines in Fig.~\ref{fig:Neff-examples}, the transition
parts between the two lines are almost the same if we plot $\Delta N_{{\rm eff}}$
as a function of $|{\cal M}|^{2}$, as shown in the right panel. This
can be understood from Eq.~\eqref{eq:small-m-pre} where the varying
part is $\propto(g_{Z'}m_{Z'})^{2}$, or more fundamentally, from
Eq.~\eqref{eq:-40} which implies that the freeze-in production in
the relativistic regime depends only on $|{\cal M}|^{2}$. From the
right panel, one can also infer that larger (smaller) $m_{Z'}$ always
leads to smaller (larger) $\Delta N_{{\rm eff}}$ if $|{\cal M}|^{2}$
is fixed.

\begin{table*}
\centering

\begin{tabular}{ccc}
\toprule 
 & $\Delta N_{{\rm eff}}$ (1$\sigma$) & $\Delta N_{{\rm eff}}$ (2$\sigma$)\tabularnewline
\midrule 
Planck 2018 \cite{Planck:2018vyg} & $\Delta N_{{\rm eff}}<0.115$ & $\Delta N_{{\rm eff}}<0.285$\tabularnewline
SO \cite{SimonsObservatory:2019qwx,SimonsObservatory:2018koc} & $\Delta N_{{\rm eff}}<0.05$ & $\Delta N_{{\rm eff}}<0.1$\tabularnewline
CMB-S4 \cite{Abazajian:2019eic,CMB-S4:2016ple} & $\Delta N_{{\rm eff}}<0.03$ & $\Delta N_{{\rm eff}}<0.06$\tabularnewline
CMB-HD \cite{CMB-HD:2022bsz} & $\Delta N_{{\rm eff}}<0.014$ & $\Delta N_{{\rm eff}}<0.028$\tabularnewline
\bottomrule
\end{tabular}

\caption{\label{tab:CMB-Neff} Current experimental bounds on $\Delta N_{{\rm eff}}$
and future sensitivity reach.}
\end{table*}

\begin{figure}
\centering

\includegraphics[width=0.9\textwidth]{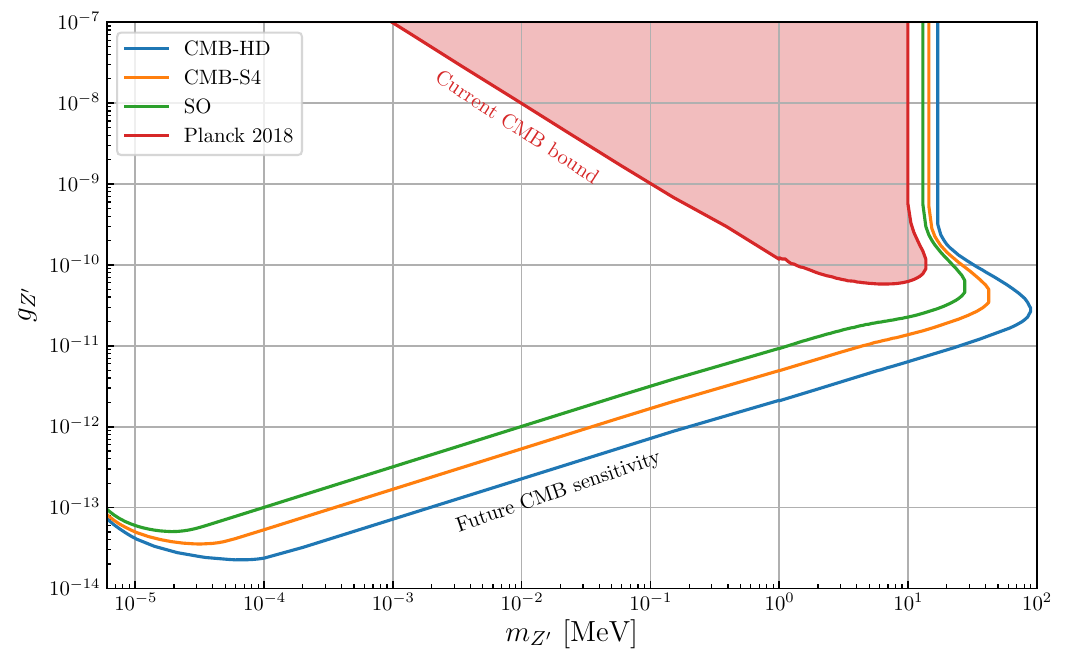}

\includegraphics[width=0.9\textwidth]{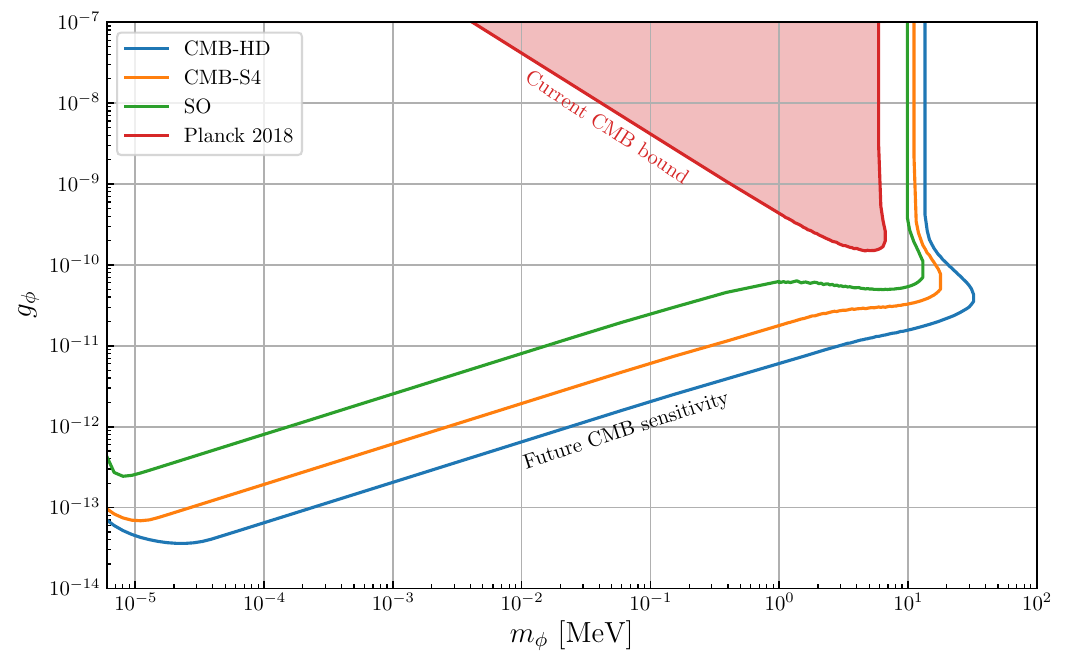}

\caption{\label{fig:main-result}Current and future experimental bounds on
light vector (upper panel) and scalar (lower panel) mediators. }
\end{figure}

Having computed the $\Delta N_{{\rm eff}}$-$g_{Z'}$ curves, we can
compare them with $\Delta N_{{\rm eff}}$ measurements and obtain
constraints on $g_{Z'}$. The latest CMB measurement of $N_{{\rm eff}}$
comes from Planck 2018~\cite{Planck:2018vyg}, which gives $N_{{\rm eff}}=2.99\pm0.17$
at $1\sigma$ C.L. After subtracting the SM value, we take $\Delta N_{{\rm eff}}<2.99+0.17\times2-3.045=0.285$
as the $2\sigma$ upper bound. For future CMB experiments, we select
three experiments, namely Simons Observatory (SO) \cite{SimonsObservatory:2019qwx,SimonsObservatory:2018koc},
CMB-S4 \cite{Abazajian:2019eic,CMB-S4:2016ple}, and CMB-HD \cite{CMB-HD:2022bsz}.
The anticipated sensitivity reach of these experiments are listed
in Tab.~\ref{tab:CMB-Neff}. The constraints on $g_{Z'}$ derived
from the experimental bounds are presented in the upper panel of Fig.~\ref{fig:main-result},
together with the constraints on $g_{\phi}$ for the scalar case in
the lower panel. 

As is manifest in Fig.~\ref{fig:main-result}, there would be substantial
improvement of the future sensitivity reach in the sub-MeV region
by orders of magnitude. The crucial difference between current and
future experiments is that the former cannot reach the Post-$\nu$DEC
equilibrium value of $\Delta N_{{\rm eff}}$. The Planck 2018 measurement
of $N_{{\rm eff}}$ is only able to probe the area between the Post-
and Pre-$\nu$DEC equilibrium lines in Fig.~\ref{fig:Neff-examples},
while future measurements can readily  reach the area below the Post-$\nu$DEC
equilibrium and thus probe the dilution-resistant effect. 

The result presented here, therefore, implies the great significance
of further improving the measurement of $N_{{\rm eff}}$. Even if
the current bound on $\Delta N_{{\rm eff}}$ is improved by a factor
of two, an enormously large part of the unexplored parameter space
will be unveiled. 

There are several noteworthy features of the bounds presented in Fig.~\ref{fig:main-result},
to be discussed below. 

First, the current CMB bound (Planck 2018) on $g_{Z'/\phi}$ goes
up as $m_{Z'/\phi}$ decreases, but the product $g_{Z'/\phi}m_{Z'/\phi}$
is roughly a constant. This corresponds to the overlapping part of
the sub-MeV curves in the right panel of Fig.~\ref{fig:Neff-examples}.
As previously discussed, this essentially stems from that the freeze-in
production in the relativistic regime depends only on $|{\cal M}|^{2}\propto(g_{Z'/\phi}m_{Z'/\phi})^{2}$. 

Second, for some large $m_{Z'/\phi}$, there are both upper and lower
bounds on $g_{Z'/\phi}$. This is due to the non-monotonic behavior
shown in the right panel of Fig.~\ref{fig:approx-Neff} and it can
be understood from Eq.~\eqref{eq:large-m-pre}. At masses around $10$
MeV, the bounds become vertical, implying that $\Delta N_{{\rm eff}}$
is independent of the couplings in this regime. It corresponds to
the scenario that $Z'$/$\phi$ is in thermal equilibrium with the
SM plasma before neutrino decoupling. Although the number density
of $Z'$ or $\phi$ at $T=T_{\nu}^{{\rm dec}}$ is Boltzmann suppressed,
 it still carries a non-negligible amount of energy at neutrino decoupling.
This part of energy will eventually be injected to the neutrino sector
and increase $N_{{\rm eff}}$. So $\Delta N_{{\rm eff}}$ in this
regime depends only on the mass. 

Finally, the curves for future sensitivity below about 10 eV stops
decreasing as $m_{Z'/\phi}$ decreases. This is because in our code,
we cut the evolution at $T_{\gamma}=1$ eV while  it is possible that
very light $Z'$/$\phi$ have not fully decayed at this point. Below
$T_{\gamma}=1$ eV, the evolution would encounter the radiation-matter
equality (at $T_{\gamma}\approx0.8$ eV) and the recombination (at
$T_{\gamma}\approx0.25$ eV). Furthermore, neutrino masses would be
non-negligible at the sub-eV scale. By cutting the evolution at $T_{\gamma}=1$
eV and taking neutrino energy only for $N_{{\rm eff}}$ evaluation,
we obtain relatively conservative bounds in the lower left corners
of these plots.  A more dedicated study on eV-scale $Z'$/$\phi$
is presented in Ref.~\cite{Sandner:2023ptm}, which shows that in
addition to the effect caused by the energy stored in $Z'$/$\phi$,
there is another important effect related to neutrino free streaming.
Including this effect can substantially improve the current CMB bound
on eV-scale $Z'$/$\phi$. 

\begin{figure}
\centering

\includegraphics[width=0.7\textwidth]{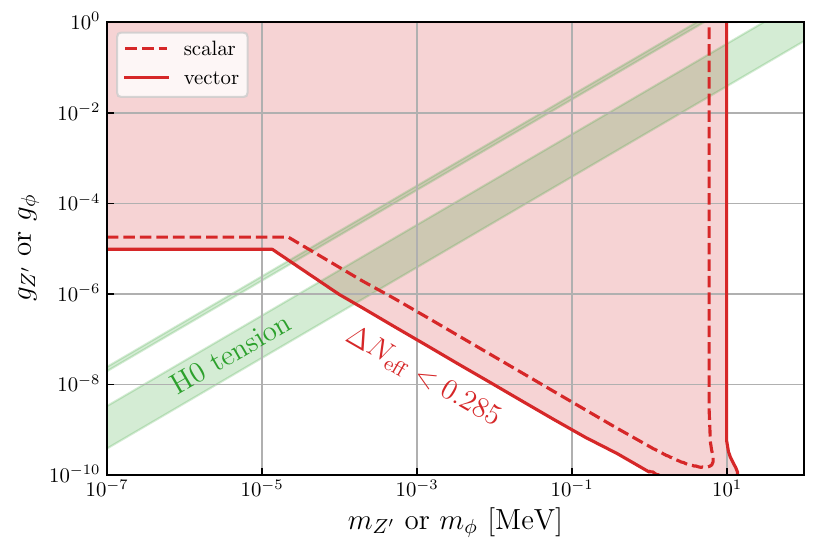}

\caption{$N_{{\rm eff}}$ constraints on neutrinophilic light mediators compared
with the neutrino self-interactions favored by the $H_{0}$ tension.
\label{fig:nuself}}
\end{figure}

Recently, there has been a notable surge of interest in neutrino
self-interactions, partly sparked by the Hubble tension---see e.g.~\cite{Blinov:2019gcj}
and references therein. Neutrino self-interactions have been proposed
in Ref.~\cite{Kreisch:2019yzn} as one of the possible solutions
to the tension.  The suggested strength of neutrino self-interactions
to resolve the Hubble tension, in terms of four-fermion contact interactions,
is 
\begin{equation}
G_{\nu\text{SI}}/G_{F}\in[3.22\times10^{9},\ 5.05\times10^{9}]\oplus[1.3\times10^{6},\ 1.1\times10^{8}]\thinspace,\label{eq:-42}
\end{equation}
where $G_{F}$ is the SM fermion constant and $G_{\nu\text{SI}}$
denotes the strength of neutrino self-interactions. This is many orders
of magnitude stronger than the SM prediction ($\sim G_{F}$). Yet
it still cannot be fully excluded after considering various constraints~\cite{Berryman:2018ogk,Blinov:2019gcj,Brdar:2020nbj,Deppisch:2020sqh}.
 Here, we demonstrate that precision measurements of $N_{{\rm eff}}$
can impose stringent constraints on such strong neutrino self-interactions.
Since $G_{\nu\text{SI}}^{-1/2}$ is around $1\sim100$ MeV, the four-fermion
interaction has to be opened up at energy scales above $G_{\nu\text{SI}}^{-1/2}$
and thus implies new mediators lighter than $\sim100$ MeV, being
it either a scalar ($\phi$) or vector ($Z'$). In Fig.~\ref{fig:nuself},
the green bands show the required $g_{Z'/\phi}$ and $m_{Z'/\phi}$
to produce the interaction strength in Eq.~\eqref{eq:-42}. The current
$N_{{\rm eff}}$ bounds are plotted as the red solid and dashed lines
for vector and scalar mediators, respectively. For $m_{Z'/\phi}$
below $10^{-5}$ MeV, the bounds are flat because we have included
$\nu\overline{\nu}\leftrightarrow2Z'$ and $\nu\overline{\nu}\leftrightarrow2\phi$,
which could effectively thermalize $Z'$/$\phi$ before neutrino decoupling
if the couplings are above $\sim10^{-5}$. 

\begin{figure}
\centering

\includegraphics[width=0.99\textwidth]{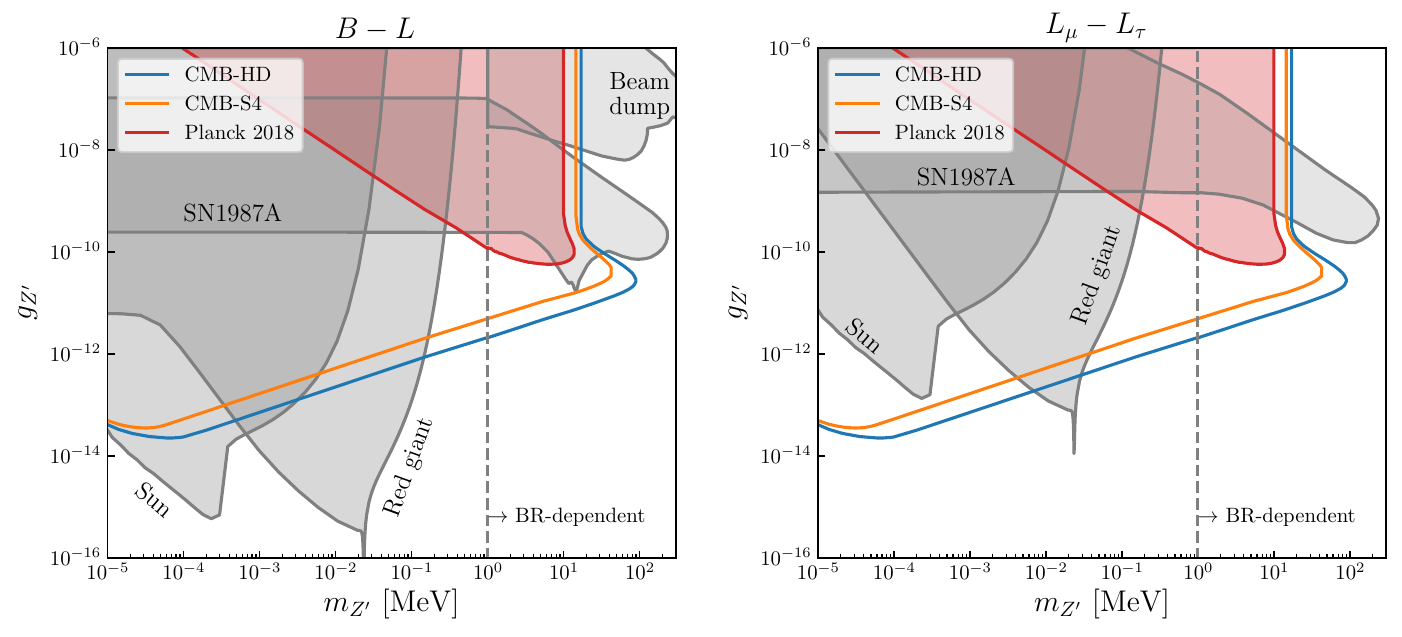}

\caption{$N_{{\rm eff}}$ constraints on light $Z'$ in the $B-L$ and $L_{\mu}-L_{\tau}$
models compared with other known bounds derived from Supernova 1987A
(SN1987A)~\cite{Croon:2020lrf},  stellar cooling of the Sun and
red giants~\cite{Li:2023vpv}, and beam dump experiments~\cite{Bauer:2018onh}.
The vertical dashed lines indicate that above 1 MeV, the $N_{{\rm eff}}$
constraints actually depend on the branching ratio of $Z'$ decay,
which is not included in the presented results. \label{fig:model}}
\end{figure}

Our results can also be applied to some specific models of light mediators
which are coupled to other SM fermions as well. For demonstration,
we select the $B-L$ and $L_{\mu}-L_{\tau}$ models and plot the $N_{{\rm eff}}$
constraints together with other relevant bounds in Fig.~\ref{fig:model}.
It is known that astrophysical bounds on light $Z'$ are stringent.
Here we take supernova bounds (labeled SN1987A) and stellar cooling
bounds from Refs.~\cite{Croon:2020lrf,Li:2023vpv}. Within the shown
window in Fig.~\ref{fig:model}, most laboratory bounds are irrelevant
except for the beam dump bound. For $B-L$, we take the beam dump
bound from Ref.~\cite{Bauer:2018onh} and add it to the left panel
of Fig.~\ref{fig:model}. The beam dump bound is invalid for $m_{Z'}\lesssim1$
MeV because such light $Z'$ cannot decay to electrons. For $L_{\mu}-L_{\tau}$,
the beam dump bound is absent because the dominant channels of $Z'$
below $\sim200$ MeV are invisible ($Z'\to\nu_{\mu}\overline{\nu_{\mu}}$,
$\nu_{\tau}\overline{\nu_{\tau}}$). Future experiments like SHiP
might have sensitivity to such a $Z'$~\cite{Coy:2021wfs}. In addition
to the beam dump bounds, the next potentially important bounds are
from neutrino-electron scattering (e.g. Borexino, CHARM II, TEXONO)~\cite{Bauer:2018onh,Lindner:2018kjo}.
But they typically set upper limits on the couplings above $10^{-6}$,
so they are absent in Fig.~\ref{fig:model}. 

It is important to note that, below 1 MeV, the $Z'$ in $B-L$ and
$L_{\mu}-L_{\tau}$ can only decay to neutrinos. For $g_{Z'}$ below
$10^{-6}$, the new interactions between neutrinos and electrons mediated
by $Z'$ cannot significantly modify neutrino decoupling. Therefore,
for $Z'$ below 1 MeV, our results are not significantly affected
by $Z'$-electron interactions. For $Z'$ in the $1\sim100$ MeV range,
then the results depend on the branching ratio of $Z'$ decay. More
specifically, it depends of the ratio of $Z'\to e^{+}e^{-}$ to $Z'\to\nu\overline{\nu}$.
In the $B-L$ model, the latter dominates over the former due to three
neutrino flavors. In the $L_{\mu}-L_{\tau}$ model, $Z'\to e^{+}e^{-}$
is possible only at the one-loop level and hence suppressed. So even
though we have not included the branching ratio in the $N_{{\rm eff}}$
constraints presented here, we anticipate that the branching ratio
would only lead to limited changes of our results.

\section{Conclusions \label{sec:Conclusions}}

The effective relativistic neutrino species, $N_{{\rm eff}}$, is
sensitive to new light particles that carry significant energy in
the early universe after neutrinos decoupled. In this work, we investigate
$N_{{\rm eff}}$ constraints on a light mediator, which can be either
a scalar ($\phi$) or a vector ($Z'$), that is primarily coupled
to neutrinos. The main results are presented in Fig.~\ref{fig:main-result}
and well understood from our analytical estimates elaborated in Sec.~\ref{sec:Analytical}. 

For $\phi/Z'$ heavier than a few MeV, most of the $\phi/Z'$ particles
are produced before neutrino decoupling so their contributions to
$N_{{\rm eff}}$ mainly depend on the amount of the energy carried
by $\phi/Z'$ into the decoupled neutrino sector.  In this regime,
current bounds and the sensitivity reach of future experiments are
qualitatively similar. 

For $\phi/Z'$ much lighter than the MeV scale, however, future CMB
experiments such as CMB-S4, SO, CMB-HD can substantially improve the
current limits on the couplings of $\phi/Z'$ by many order of magnitude.
This is because very light $\phi/Z'$ with sufficiently weak couplings
can only be produced at very low temperatures. If they are predominantly
produced from decoupled neutrinos, their contributions to $N_{{\rm eff}}$
are only generated by the dilution-resistant effect---see e.g. Eq.~\eqref{eq:-4}.
The current measurements of $N_{{\rm eff}}$ do not have sufficient
precision to probe the dilution-resistant effect, since its maximal
contribution to $N_{{\rm eff}}$ is $0.118$ and $0.242$ (see Tab.~\ref{tab:t}
and Fig.~\ref{fig:Neff-examples}) for scalar and vector, respectively.
These values can be readily reached by the aforementioned future experiments. 

Therefore, our results imply the great significance of further improving
the measurement of $N_{{\rm eff}}$. With modest improvements foreseeable
in the next-generation experiments, an enormously large part of the
unexplored parameter space of new particles coupled to neutrinos will
be unveiled. 

\appendix

\section{Collision terms\label{sec:Collision}}

When the coupling $g_{Z'}$ is sufficiently small, the dominant processes
for $Z'$ production and depletion is $\nu\overline{\nu}\to Z'$ and
$Z'\to\nu\overline{\nu}$. In our convention, $n$ and $\rho$ do
not include the internal degrees of freedom. To avoid potential confusions
in dealing with the internal degrees of freedom such as $N_{\nu}=3$
for three neutrino flavors and $N_{Z'}=3$ for three polarizations
of $Z'$, we write down the dependence on $N_{Z'}$ and $N_{\nu}$
explicitly:
\begin{align}
C_{{\rm prod.}}^{(n_{Z'})}=N_{\nu}C_{\nu\overline{\nu}\to Z'}^{(n_{Z'})}\thinspace,\ \  & C_{{\rm depl.}}^{(n_{Z'})}=N_{\nu}C_{Z'\to\nu\overline{\nu}}^{(n_{Z'})}\thinspace,\label{eq:-43}\\
C_{{\rm prod.}}^{(n_{\nu})}=N_{Z'}C_{Z'\to\nu\overline{\nu}}^{(n_{\nu})}\thinspace,\ \  & C_{{\rm depl.}}^{(n_{\nu})}=N_{Z'}C_{\nu\overline{\nu}\to Z'}^{(n_{\nu})}\thinspace,\label{eq:-44}\\
C_{{\rm prod.}}^{(\rho_{Z'})}=N_{\nu}C_{\nu\overline{\nu}\to Z'}^{(\rho_{Z'})}\thinspace,\ \  & C_{{\rm depl.}}^{(\rho_{Z'})}=N_{\nu}C_{Z'\to\nu\overline{\nu}}^{(\rho_{Z'})}\thinspace,\label{eq:-45}\\
C_{{\rm prod.}}^{(\rho_{\nu})}=N_{Z'}C_{Z'\to\nu\overline{\nu}}^{(\rho_{\nu})}\thinspace,\ \  & C_{{\rm depl.}}^{(\rho_{\nu})}=N_{Z'}C_{\nu\overline{\nu}\to Z'}^{(\rho_{\nu})}\thinspace.\label{eq:-46}
\end{align}
For the scalar case, we simply need to replace $N_{Z'}\to N_{\phi}=1$. 

For a generic process $12\to3$, let us denote the momentum, energy,
and phase space distribution of the $i$-th particle by $\mathbf{p}_{i}$,
$E_{i}$, and $f_{i}$, respectively. Then the collision terms are
formulated as follows:
\begin{align}
C_{12\to3}^{(n_{3})} & =\int d\Pi_{1}d\Pi_{2}d\Pi_{3}f_{1}f_{2}(1\mp f_{3})|{\cal M}|^{2}(2\pi)^{4}\delta^{4}\thinspace,\label{eq:-47}\\
C_{12\to3}^{(n_{1})} & =\int d\Pi_{1}d\Pi_{2}d\Pi_{3}f_{1}f_{2}(1\mp f_{3})|{\cal M}|^{2}(2\pi)^{4}\delta^{4}\thinspace,\label{eq:-48}\\
C_{12\to3}^{(\rho_{3})} & =\int d\Pi_{1}d\Pi_{2}d\Pi_{3}f_{1}f_{2}(1\mp f_{3})E_{3}|{\cal M}|^{2}(2\pi)^{4}\delta^{4}\thinspace,\label{eq:-49}\\
C_{12\to3}^{(\rho_{1})} & =\int d\Pi_{1}d\Pi_{2}d\Pi_{3}f_{1}f_{2}(1\mp f_{3})E_{1}|{\cal M}|^{2}(2\pi)^{4}\delta^{4}\thinspace,\label{eq:-50}
\end{align}
where $d\Pi_{i}\equiv\frac{d^{3}\mathbf{p}_{i}}{2E_{i}(2\pi)^{3}}$,
$|{\cal M}|^{2}$ denotes the squared amplitude of the process, and
$\delta^{4}$ is short for $\delta^{4}(p_{1}+p_{2}-p_{3})$. Note
that $C_{12\to3}^{(n_{3})}=C_{12\to3}^{(n_{1})}$ while $C_{12\to3}^{(\rho_{3})}\neq C_{12\to3}^{(\rho_{1})}$.
In fact, using energy conservation, $E_{3}=E_{1}+E_{2}$, it is straightforward
to obtain
\begin{equation}
C_{12\to3}^{(\rho_{1})}=\frac{1}{2}C_{12\to3}^{(\rho_{3})}\thinspace.\label{eq:-51}
\end{equation}

In the Boltzmann approximation, with some substitutions such as $e^{-E_{1}/T}e^{-E_{2}/T}=e^{-E_{3}/T}$
and Eq.~\eqref{eq:-51}, all the collision terms in Eqs.~\eqref{eq:-47}-~\eqref{eq:-50}
can be reduced to the integrals computed in Appendix~A of Ref.~\cite{Luo:2020fdt}.
Then it is straightforward to obtain the collision terms:
\begin{align}
C_{\nu\overline{\nu}\to Z'}^{(n_{Z'})} & =C_{\nu\overline{\nu}\to Z'}^{(n_{\nu})}=\frac{|{\cal M}|^{2}}{32\pi^{3}}m_{Z'}T_{\nu}K_{1}\left(\frac{m_{Z'}}{T_{\nu}}\right)e^{2\mu_{\nu}/T_{\nu}}\thinspace,\label{eq:-52}\\
C_{Z'\to\nu\overline{\nu}}^{(n_{Z'})} & =C_{Z'\to\nu\overline{\nu}}^{(n_{\nu})}=\frac{|{\cal M}|^{2}}{32\pi^{3}}m_{Z'}T_{Z'}K_{1}\left(\frac{m_{Z'}}{T_{Z'}}\right)e^{\mu_{Z'}/T_{Z'}}\thinspace.\label{eq:-53}\\
C_{\nu\overline{\nu}\to Z'}^{(\rho_{Z'})} & =2C_{\nu\overline{\nu}\to Z'}^{(\rho_{\nu})}=\frac{|{\cal M}|^{2}}{32\pi^{3}}m_{Z'}^{2}T_{\nu}K_{2}\left(\frac{m_{Z'}}{T_{\nu}}\right)e^{2\mu_{\nu}/T_{\nu}}\thinspace,\label{eq:-54}\\
C_{Z'\to\nu\overline{\nu}}^{(\rho_{Z'})} & =2C_{Z'\to\nu\overline{\nu}}^{(\rho_{\nu})}=\frac{|{\cal M}|^{2}}{32\pi^{3}}m_{Z'}^{2}T_{Z'}K_{2}\left(\frac{m_{Z'}}{T_{Z'}}\right)e^{\mu_{Z'}/T_{Z'}}\thinspace.\label{eq:-55}
\end{align}
where $\mu_{\nu}$ and $\mu_{Z'}$ denote the chemical potentials
of $\nu$ and $Z'$. 

There is a subtlety regarding the scalar case: the process $\phi\to\nu\nu$
has two  identical particles in the final state so in principle one
should include a symmetry factor of $2$ in the phase space integrals.
Meanwhile, every time when the reaction happens, it produces two $\nu$,
i.e. the process is twice efficient as $Z'\to\nu\overline{\nu}$ in
producing $\nu$. One can check that overall, these factors of two
cancel out so that Eqs.~\eqref{eq:-52}-\eqref{eq:-55} can also be
applied to the scalar case with trivial substitutions such as $m_{Z'}\to m_{\phi}$
and $T_{Z'}\to T_{\phi}$. 

Since the number density of $Z'$ in the Maxwell-Boltzmann distribution
is given by 
\begin{equation}
n_{Z'}\approx\frac{1}{2\pi^{2}}m_{Z'}^{2}T_{Z'}e^{\mu_{Z'}/T_{Z'}}K_{2}\left(m_{Z'}/T_{Z'}\right),\label{eq:-56}
\end{equation}
the non-relativistic limit of Eqs.~\eqref{eq:-52}-\eqref{eq:-55}
can be written as 
\begin{align}
\lim_{m_{Z'}\to\infty}C_{\nu\overline{\nu}\to Z'}^{(n_{Z'})} & =\lim_{m_{Z'}\to\infty}C_{\nu\overline{\nu}\to Z'}^{(n_{\nu})}\approx\overline{\Gamma}_{Z'}n_{Z'}^{({\rm eq})}\thinspace,\label{eq:-57}\\
\lim_{m_{Z'}\to\infty}C_{Z'\to\nu\overline{\nu}}^{(n_{Z'})} & =\lim_{m_{Z'}\to\infty}C_{Z'\to\nu\overline{\nu}}^{(n_{\nu})}\approx\overline{\Gamma}_{Z'}n_{Z'}\thinspace,\label{eq:-58}\\
\lim_{m_{Z'}\to\infty}C_{\nu\overline{\nu}\to Z'}^{(\rho_{Z'})} & =\lim_{m_{Z'}\to\infty}2C_{\nu\overline{\nu}\to Z'}^{(\rho_{\nu})}\approx m_{Z'}\overline{\Gamma}_{Z'}n_{Z'}^{({\rm eq})}\thinspace,\label{eq:-59}\\
\lim_{m_{Z'}\to\infty}C_{Z'\to\nu\overline{\nu}}^{(\rho_{Z'})} & =\lim_{m_{Z'}\to\infty}2C_{Z'\to\nu\overline{\nu}}^{(\rho_{\nu})}\approx m_{Z'}\overline{\Gamma}_{Z'}n_{Z'}\thinspace,\label{eq:-60}
\end{align}
where we have used $\lim_{x\to\infty}K_{1}\left(x\right)\approx K_{2}\left(x\right)$
and $n_{Z'}^{({\rm eq})}$ is defined as the equilibrium value of
$n_{Z'}$. The decay width $\overline{\Gamma}_{Z'}$ does not include
$N_{Z'}$:
\begin{equation}
\overline{\Gamma}_{Z'}\ensuremath{\equiv}\frac{|{\cal M}|^{2}}{16\pi m_{Z'}}\thinspace.\label{eq:-61}
\end{equation}
For the scalar case, we have similar formulae. The potential difference
due to degrees of freedom has been fully absorbed by $N_{Z'}$ and
$N_{\phi}$. 

\section{Technical details of numerical calculations\label{sec:Technical}}

There are a few technical issues in solving the Boltzmann equation
numerically. First, the Boltzmann equation often exhibits \emph{stiffness},
which is a well-known phenomenon of ordinary differential equations
(ODE), causing numerical instability when the collision rates are
too high. In general, implicit methods such as backward differentiation
formula (BDF) are more suitable to deal with stiffness than explicit
methods. In practice, we do find that the BDF method implemented in
{\tt scipy} (e.g. the {\tt solve\_ivp} ODE solver in {\tt scipy})
is more robust than other methods against the instability. However,
in some cases when the collision terms are many orders of magnitude
higher than the Hubble expansion and meanwhile the Boltzmann suppression
starts to play a crucial role, using the BDF method still cannot generate
numerically stable solutions. 

Our approach to deal with the numerical instability in this regime
is as follows. When the collision rates are too high compared to the
Hubble expansion, e.g. $C_{{\rm prod.}/\text{depl.}}^{(n)}>10^{3}Hn$,
then the system should be in equilibrium and the left- and right-hand
sides of the reaction processes are strongly coupled by the collision
terms. For processes in the strongly-coupled regime, the actual values
of reaction rates are not important as long as they are able to maintain
the equilibrium. This implies that one can reduce the reaction rates
manually to avoid the ODE being too stiff, and meanwhile still keep
the accuracy of the solution. In practice, since the $C$ terms in
Eqs.~\eqref{eq:na3} and \eqref{eq:rhoa4} are always divided by $H$,
reducing those $C$'s are equivalent to increasing $H$. Therefore,
our actual measure to reduce the stiffness of the equation is not
to change the collision terms, but to increase $H$ in the following
smooth way\footnote{Abrupt changes in numerical functions that are fed to the ODE solver
often increases the instability due to unnecessary oscillations of
the solutions caused by the abrupt changes. For example, we have tested
that if Eq.~\eqref{eq:-62} is changed to $H=\max(H,\ \lambda\Gamma)$
and $\Gamma\equiv\max\left(C_{{\rm prod.}}^{(n)},\ C_{{\rm depl}.}^{(n)}\right)/n$
which is conceptually simpler, the numerical solutions become more
unstable. \label{fn:Abrupt}}:
\begin{equation}
H\to\sqrt{H^{2}+\lambda^{2}\Gamma^{2}}\thinspace,\ \ \Gamma\equiv\frac{1}{n}\sqrt{\left(C_{{\rm prod.}}^{(n)}\right)^{2}+\left(C_{{\rm depl}.}^{(n)}\right)^{2}}\thinspace,\label{eq:-62}
\end{equation}
where $\lambda=10^{-3}$. Eq.~\eqref{eq:-62} implies that when $H\gg10^{-3}\Gamma$,
 $H$ is almost unmodified; when $H\gg10^{3}\Gamma$, $H$ would
be pulled up to $\lambda\Gamma$ which is still well below $\Gamma$.
Physically, it corresponds to speeding up the expansion of the universe
while still keeping the reaction in equilibrium. In practice, we find
that using Eq.~\eqref{eq:-62} hardly causes visible deviations from
the true solution but substantially increases the stability. Only
when $\lambda$ is increased to ${\cal O}(0.3)$, we observe slight
deviations. 

There is an alternative treatment regarding the issue of stiffness.
When $\Gamma\gg H$, one can simply assume that  the system is in
equilibrium and use thermodynamic formulae in equilibrium for the
rest of the evolution without solving the Boltzmann equation. This
treatment is simpler (though not adopted in our code), but one should
be careful about potential  out-of-equilibrium issues as the system
further evolves. One of the well-known examples is the freeze-out
mechanism, which can drive a thermal species out of equilibrium after
the Boltzmann suppression becomes significant. For the collision terms
considered in this work, this is not of concern because when the number
density $n$ in Eq.~\eqref{eq:-1} is approaching (or slightly deviating)\footnote{For $2\leftrightarrow1$ processes, if $n$ has reached its equilibrium
value and is experiencing the Boltzmann suppression, there is actually
a small deviation of $n$ from the equilibrium value in the subsequent
evolution. The magnitude of the deviation depends on the strength
of the couplings, stronger couplings leading to smaller deviations.
Eventually, the deviation will be vanishingly small at sufficiently
low temperatures. } its equilibrium value $n^{({\rm eq})}$, Eq.~\eqref{eq:-1} can be
roughly written as $dn/dt+3Hn\propto n-n^{({\rm eq})}$, in contrast
to the two--to-two-scattering processes in the freeze-out mechanism
where we have $dn/dt+3Hn\propto n^{2}-\left(n^{({\rm eq})}\right)^{2}$.
The crucial difference between $\delta n\equiv n-n^{({\rm eq})}$
and $n^{2}-\left(n^{({\rm eq})}\right)^{2}\approx2n\delta n$ is that
the extra power of $n$ in the latter causes additional Boltzmann
suppression in the non-relativistic regime and thus leads to the freeze-out
mechanism. 

Another technical issue we have encountered concerns the use of Bessel
functions $K_{1}(x)$ and $K_{2}(x)$  in the large $x$ limit (corresponding
to the ultra non-relativistic regime). Since $\lim_{x\to\infty}K_{1}(x)=\lim_{x\to\infty}K_{2}(x)=\sqrt{\frac{\pi}{2x}}e^{-x}$,
large $x\gtrsim700$ can lead to overflow of 64-bit floating-point
numbers which are widely used in numerical calculations. In our code,
when $x=m/T$ is above 550, the numerical evaluation of $K_{1,2}$
together with other functions with exponential behaviors is switched
to analytical expressions obtained by expanding them in terms of $1/x$.
Taking the leading order of the expansion is sufficient to limit the
error below $0.1\%$. However, as mentioned in footnote~\ref{fn:Abrupt},
if the transition is not smooth, it would cause additional instability.
So in our code, we expand these functions to the fourth power of $1/x$
to reduce the instability to an acceptable level. 

\begin{acknowledgments}
We would like to thank Miguel Escudero and Bingrong Yu for useful
discussions and Junyu Zhu for assistance in data processing. This
work is supported in part by the National Natural Science Foundation
of China under grant No. 12141501 and also by CAS Project for Young Scientists in Basic Research (YSBR-099).
\end{acknowledgments}

\bibliographystyle{JHEP}
\bibliography{ref}

\end{document}